\documentclass[twocolumn,secnumarabic,amssymb, aps, prl]{revtex4-1}
\usepackage{amsmath, amssymb}

\usepackage{graphicx}
\usepackage{color}
\usepackage{stackengine,graphicx}
\usepackage{textcmds}
\usepackage{mathtools}
\usepackage{mathtools}

\usepackage{makecell}

\newcommand{\vbr}{\textbf{r}}
\newcommand{\epow}[1]{\mathrm{e}^{#1}}

\begin{document}

\title{Collective behavior of active molecules: dynamic clusters, holes and active fractalytes}

\author{Sebastian Fehlinger}%
\affiliation{Technische Universit{\"a}t Darmstadt}

\author{Benno Liebchen}%
\email[]{benno.liebchen@pkm.tu-darmstadt.de}
\affiliation{Technische Universit{\"a}t Darmstadt}
\date{July 2023}%

\begin{abstract}
Recent experiments have led to active colloidal molecules which aggregate from non-motile building blocks and acquire self-propulsion through their non-reciprocal interactions. Here, we model the collective behavior of such active molecules and predict, besides dynamic clusters, the existence of a so-far unknown state of active matter made of `active fractalytes' which are motile crystallites featuring internal holes, gaps and a fractal dimension. These structures could serve as a starting point for the creation of active materials with a low density and mechanical properties that can be designed through their fractal internal structure.
\end{abstract}

\maketitle

\textit{Introduction.}---A general theme which is shared by all fundamental forces in nature, from the strong force acting at the scale of elementary particles to gravity at intergalactical scales is the action-reaction symmetry. This symmetry, considered as the `cornerstone of physics' by Ernst Mach \cite{Mach-01}, is at the heart of the conservation laws for momentum and angular momentum expressing the homogeneity and isotropy of space itself. If action-reaction symmetry was broken, matter could spontaneously acquire a center of mass motion, provoking an obvious conflict with conservation laws.
Accordingly, in equilibrium systems, the action-reaction symmetry applies also to effective interactions, from the interactions among nucleons in the core of all atoms to the broad variety of screened electrostatic and entropy-driven effective interactions governing the behavior of soft matter and biosystems. Interestingly, however, in non-equilibrium systems, effective interactions are in principle allowed to break action-reaction symmetry \cite{Ivlev-PRX-2015} through a subtle coupling of the system to its environment, which pays the energy and momentum bill for the spontaneous emergence of a directed motion \cite{Marchetti-RMP-2013, Elgeti-RPP-2015}.

To imagine a specific mechanism leading to nonreciprocal interactions, consider two spherical colloidal particles, one of which is isotropically coated with a catalyst that provokes a certain chemical reaction in the surrounding solvent. While the resulting concentration profile is essentially isotropic with respect to the catalytic colloid, the second colloid experiences a concentration gradient across its surface. This gradient drives a directed flow within the colloids interfacial layer by diffusioosmosis or a similar mechanism \cite{Schmidt-JCP-2019, Reigh-SM-2018, Stuermer-JCP-2019}. Because momentum is conserved in the overall system, the second colloid ballistically moves opposite to the osmatic solvent flow across its surface \cite{Schmidt-JCP-2019}, while the catalytically coated colloid essentially rests. Remarkably, even if the second colloid approaches the catalytic one and both colloids are in close contact to form a `colloidal' dimer \cite{Singh-AdMat-2017, Hauke-JCP-2020, Schmidt-JCP-2019}, the simplest realization of an active colloidal molecule \cite{Loewen-EPL-2018}, they persistently move ballistically as long as chemical reactions take place.

Recent years have made it possible to realize such ideas based on colloidal microparticles creating gradients e.g. in concentration \cite{Howse-PRL-2007, Golestanian-PRL-2005, Golestanian-NewJPhys-2007, Wuerger-PRL-2015, Rueckner-PRL-2007}, electric potential \cite{Moran-PRE-2010, Paxton-JACS-2004, Paxton-ChemEurJ-2005} or temperature \cite{Rings-PRL-2010,Gaspard-JStatMech-2019, Jiang-PRL-2010, Wuerger-PRL-2007, Auschra-EPJ-2021, Fraenzl-NatComm-2022, Braun-ACS-2013} inducing nonreciprocal attractions of tracer particles. In particular, recent experiments have realized binary mixtures of nonmotile isotropic colloids, which are driven by light \cite{Schmidt-JCP-2019, Grauer-NatCom-2021} or ion-exchange processes \cite{Niu-ACS-2018} and exhibit nonreciprocal interactions among an `attractive' and a `responsive' particle species, which lead to directed motion of the emerging active molecules. While many theoretical \cite{Soto-PRE-2015, Soto-PRL-2014, Vuijk-PRE-2022, Gonzales-NewJPhys-2019, Varma-SM-2018} and experimental \cite{Schmidt-JCP-2019, Grauer-NatCom-2021, Niu-ACS-2018, Wang-NatComm-2020, Niu-PRL-2017} works on active colloidal molecules have focused on the formation, characterization and early-stage clustering of active molecules, theoretical works \cite{Saha-PRX-2020, Fruchart-Nat-2021, Agudo-Canalejo-PRL-2019, Kreienkamp-NewJPhys-2022, Loos-NewJPhys-2020, Zjihong-PNAS-2020, Loos-arxiv-2022} have recently started to also predict the possible large scale behavior of many nonreciprocally interacting particles.  

Here, we explore a model which is directly inspired by recent experiments \cite{Schmidt-JCP-2019,Grauer-NatCom-2021, Niu-ACS-2018} and show that many interacting active molecules can self-organize into phases which have not yet been reported in the literature. First, for relatively weak nonreciprocal interactions and/or a small amount of attractive particles in the system, we observe that active molecules self-organize into arrested, dynamic clusters with a self-limited characteristic size. This phase resembles dynamic clustering in self-propelled Janus colloids \cite{Theurkauff-PRL-2012, Buttinoni-PRL-2013, Palacci-Science-2013, Ginot-NatComm-2018, Stark-AccChemRes-2018, Pohl-PRL-2014, Foffano-SM-2019, Liebchen-PRL-2015, Pohl-EPJ-2015, Huang-NJP-2017, Ma-AdThSim-2020}, but emerges from a different mechanism. Second, for stronger nonreciprocal interactions, we observe that individual active molecules self-assemble into macroscopic aggregates which feature a persistent dynamics and a characteristic internal structure which involves holes and gaps leading to a fractal (non-integer) dimension. Accordingly, we call these structures active fractalytes - a portmanteau of fractals and crystallites. These super structures exemplify a route towards the creation of amorphous, solid-like active materials which are lighter than active systems in glassy or crystalline states and feature a density that can be tuned by the composition of the system or by the chemical (phoretic) activity of its components. \\

\noindent{}
\begin{figure*}
 \noindent
 \begin{minipage}[]{0.99\textwidth}
 \begin{center}
  \includegraphics[width = 0.95\textwidth]{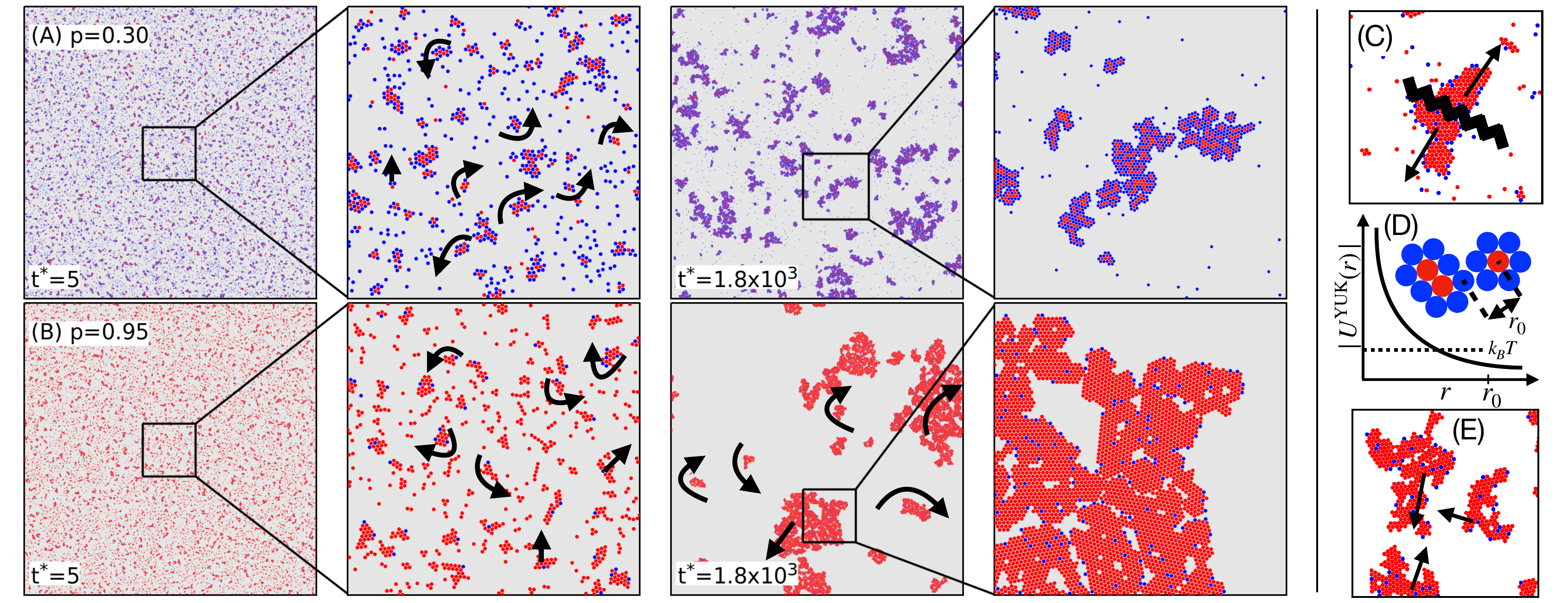}
  \end{center}
 \end{minipage}
 \caption{Snapshots from simulations starting with a random, uniform initial state at times $t^* = t D_T/\sigma^2$ showing dynamic clusters with a self-limited size for $p=0.3$ (A) and active fractalytes for $p=0.95$ (B). Black arrows indicate self-propulsion and self-rotation of the aggregates. Mechanisms leading to finite cluster sizes (C,D) and the emergence of holes within active fractalytes (E). Red particles are the attractive and blue ones the passive. Parameters: $N=2.5\times10^4, \ Y^*=300$.}
 \label{snapshots}
\end{figure*}

\textit{Model.}---Inspired by recent experiments \cite{Schmidt-JCP-2019, Grauer-NatCom-2021, Niu-ACS-2018} we define a minimal model which consists of $N=N_a+N_p$ colloidal particles in two spatial dimensions, $N_a$ of which attract all particles via an effective Yukawa potential $U_P (r) = -Y \frac{\epow{-\mu r}}{r}$, which represents attractive phoretic (far-field) interactions \cite{Soto-PRE-2015, Soto-PRL-2014,Liebchen-CK-2019, Liebchen-JCP-2019,Liebchen-JPCM-2022}. The remaining $N_p$ particles interact purely repulsively with each other. Thus, the attractive particles attract the passive particles but not vice versa, leading to nonreciprocal interactions. The overdamped Langevin equation for particle $i$ reads

\begin{equation}\label{langevin}
 \dot{\vbr}_i (t) = \frac{1}{\gamma} \textbf{F}_i + \sqrt{2D_t} \boldsymbol{\eta}_i(t), \ \ i=1, \dots, N_a, \dots, N
\end{equation}

with forces

\begin{equation}
 \textbf{F}_i = -\sum_{\genfrac{}{}{0pt}{} {j=1}{j \neq i}}^{N} \boldsymbol{\nabla}_{\textbf{r}_i} U_S(r_{ij}) - \sum_{\genfrac{}{}{0pt}{} {j=1}{j \neq i}}^{N_a} \boldsymbol{\nabla}_{\textbf{r}_i} U_P(r_{ij})
\end{equation}

where $r_{ij} = |\vbr_i - \vbr_j|$ and $\gamma$ is the Stokes drag coefficient. $D_t$ is the translational diffusion coefficient and $\boldsymbol{\eta}_i(t)$ is Gaussian white noise with zero mean and unit variance describing thermal Brownian motion.The steric repulsion is given by the Weeks-Chandler-Anderson (WCA) potential $U_S (r) = 4\epsilon \left[\left(\frac{\sigma}{r}\right)^{12} - \left(\frac{\sigma}{r}\right)^6 \right] + \epsilon$ for $r \leq 2^{1/6}\sigma$ and $U_S(r) = 0$ for $r>2^{1/6}\sigma$, where $\sigma$ denotes the (soft) particle diameter. 
To reduce the parameter space, we now use $\sigma$, $\sigma^2/D_t$ and $k_BT$ as length, time and energy units, respectively. The remaining dimensionless parameters are the packing fraction $\Phi_0 =N\pi\sigma^2/4L^2$, where $L$ is the length of the quadratic simulation box, the fraction of attractive particles $p=N_a/N$, the effective inverse  screening length $\mu^*=\mu \sigma$, the attraction strength $Y^*=Y/(k_BT\sigma)$ and the strength of the WCA potential $\epsilon^*=(\epsilon /k_BT)$ (see SM \cite{supp-info} for the reduced equations of motion). In experiments, it can be realized with nonreciprocally interacting colloidal mixtures \cite{Grauer-NatCom-2021, Hauke-JCP-2020}, with ion-exchange driven particle-tracer mixtures \cite{Niu-ACS-2018} or possibly also with nonreciprocal dusty plasmas \cite{Ivlev-PRX-2015} or oil exchanging droplets \cite{Meredith-Natchem-2020}. To be generic, we disregard system specific ingredients, such as a possible binary nature of the solvent which can in general lead to additional interesting phenomena \cite{Grauer-NatCom-2021}. Notably, as opposed to many other works exploring nonreciprocal interactions such as  \cite{Hauke-JCP-2020, Singh-AdMat-2017, Vuijk-PRE-2022, Knezevic-Nature-2022}, as in experiments \cite{Schmidt-JCP-2019, Grauer-NatCom-2021}, we consider isotropic particles which just have spatial degrees of freedom but no orientational ones, so that the individual particles in our model move like passive Brownian particles which acquire activity only through interactions with particles of the other species.

We perform Brownian dynamics simulations with up to $2.5 \times 10^4$ particles in a box of size $L \times L$ with periodic boundary conditions by using the software package LAMMPS \cite{Thompson-CPC-2022}. Here, we mainly vary the key control parameters $Y^*$ and $p$, fix $\Phi_0=0.1$ to stay at low packing fraction, $\epsilon^*=100$ so that the particles can hardly overlap and also $\mu^*=2$ so that the effective screening length of the phoretic attractions is roughly comparable to the particle size \cite{Hauke-JCP-2020, Liebchen-JCP-2019}. \\

\begin{figure*}[!t]
  \begin{center}
  \includegraphics[width = 0.99\textwidth]{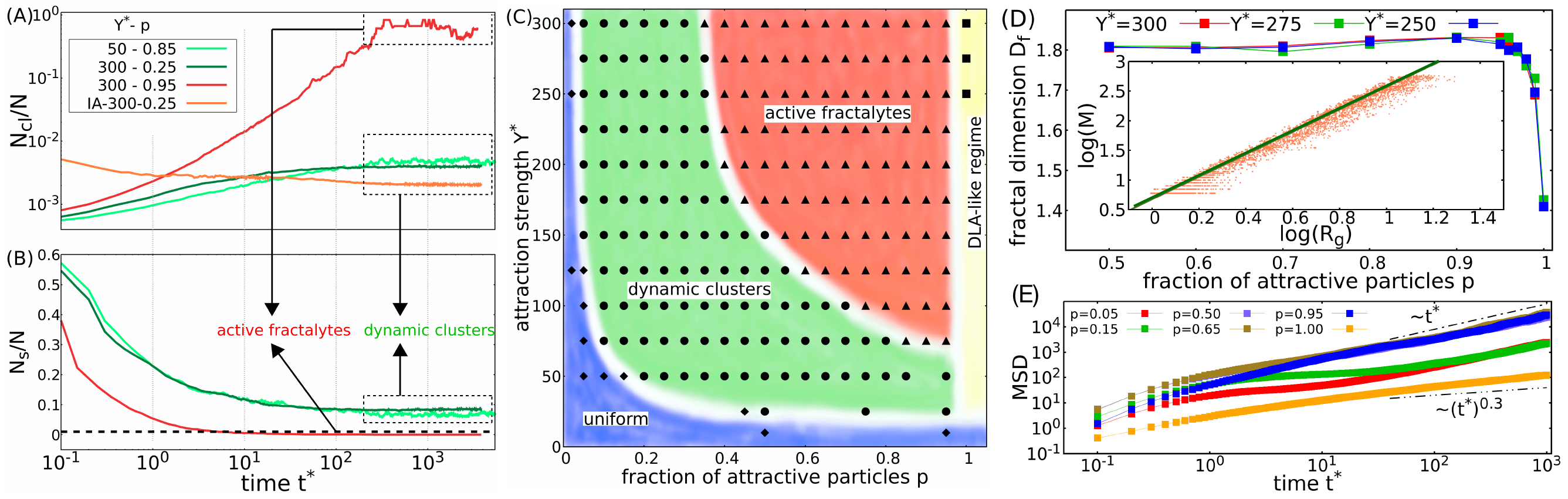}
  \end{center}
 \caption{Phase diagram, cluster sizes and fractal dimension: Typical time-evolution (moving averages) of the mean cluster size $N_{cl}(t^*)/N$ (A) and the number of single particles $N_s(t^*)/N$ (B) for parameters shown in the key, where the additional `IA' refers to aggregated initial state as shown in the SM \cite{supp-info}. The black dotted line in panel (B) indicates the threshold of $0.01N$. Non-equilibrium phase diagram (C). Panel (D) shows the dependence of the fractal dimension $D_f$ on the fraction of attractive particles $p$ for different attraction strengths $Y^*$ and the inset shows one exemplary fit of Eq. (\ref{fit-log}) to the corresponding data for $(Y^*,p)=(300,0.95)$. Mean-squared displacement (MSD) for $Y^*=300$ and different $p$ as shown in the key (E).}
 \label{phasediagram}
\end{figure*}

\begin{figure}[]
\begin{center}
\includegraphics[width=\linewidth]{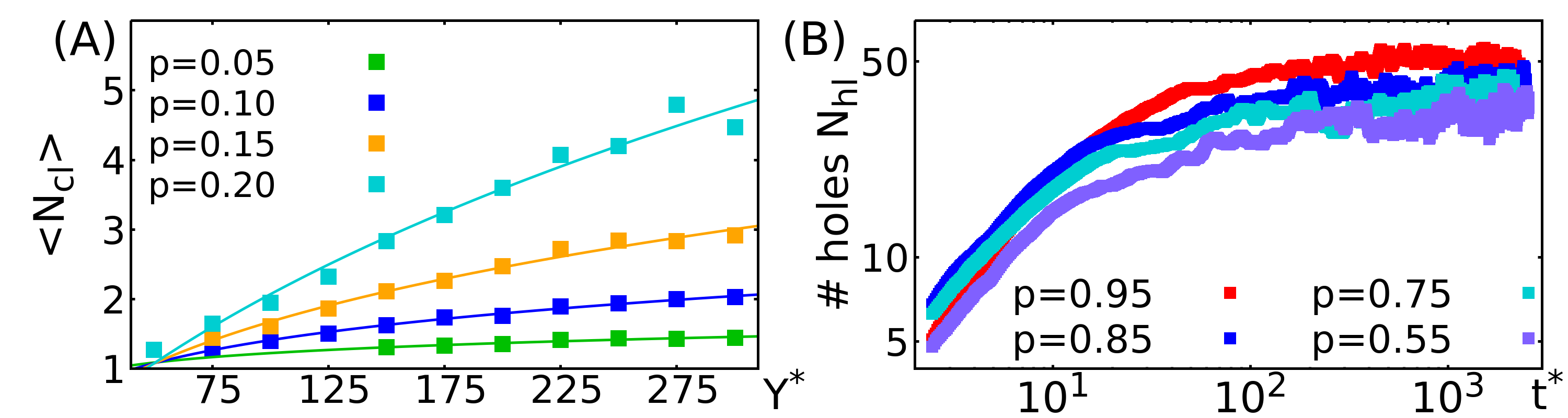}
 \end{center}
 \caption[]{(A) Characteristic cluster size $\langle N_{cl}\rangle$ as a function of the attraction strength $Y^*$ for different $p$. Lines show fits to the predicted scaling. (B) Number of holes $N_{hl}$ as a function of $t^*$.}
 \label{length-scales}
\end{figure}

\textit{Activity-induced fragmentation arrests coarsening.}---For moderately strong attractions $Y^*<20$ ( $Y^* = 10$ corresponds to an interaction strength which is comparable to the thermal energy at contact, i.e. $r^*=2^{1/6}\sigma$), independently of $p$, the system ends up in a disordered uniform state. Choosing stronger attractive interactions ($Y^* > 20$), at early times active molecules form with a huge variety in shape and dynamics, comprising ballistically moving dimers, trimers and larger molecules following chiral trajectories (Fig. \ref{snapshots}), similarly as observed in experiments \cite{Schmidt-JCP-2019, Niu-ACS-2018}. The molecules move persistently since attractive particles undirectionally attract passive ones, which push the active ones forword due to the WCA-repulsions. At later times, active molecules aggregate and successively form larger and larger clusters which coexist with a gas-like phase comprising individual non-attractive particles only (Fig. \ref{snapshots}A). Interestingly, at some time, despite the presence of attractions, the average cluster size stops to increase (green lines in Fig. \ref{phasediagram}A) and converges to a characteristic finite size. We have identified two different mechanisms which can arrest coarsening and lead to finite cluster sizes. First, if the attractions are comparatively weak (e.g. $(Y^*,p)=(50,0.85)$, Movie M1 in the SM \cite{supp-info}), thanks to nonreciprocal forces acting on their components and thermal motion clusters split up from time to time leaving behind smaller cluster fragments which self-propel into different directions (Fig. \ref{snapshots}C). This activity-induced cluster fragmentation arrests coarsening and leads to clusters with a self-limited size of $\thicksim 10$ particles (Fig. \ref{phasediagram}A, light-green line), which look similar to those observed in experiments made of individually self-propelling Janus colloids \cite{Buttinoni-PRL-2013}. In addition to cluster fragmentation processes also single particles frequently detach from existing clusters. Second, for a quite low percentage of attractive particles and a strong attraction (e.g. $(Y^*,p)=(300,0.25)$, Movie M2 in the SM \cite{supp-info}) clusters tend to posses a layer of passive particles surrounding them (Fig. \ref{snapshots}A,D), which prevents them from growing beyond a certain characteristic size (Fig. \ref{phasediagram}A, dark-green line) because the attractions are effectively screened (Yukawa potential) and do no longer dominate over thermal motion at distances of two or more particle diameters. We now ask how the characteristic cluster size depends on the attraction strength $Y^*$ and on the composition of the system $p$. For small values of $p$, where the collective dynamics of cluster fragments is rather unimportant, this length scale can be understood based on the competition between passive diffusion and attractions since clusters are expected to grow until both effects balance each other. In the SM \cite{supp-info} we determine the following relation for the mean-number of particles in a cluster at late times: $\langle N_{cl}\rangle\propto W(aY^*)$. Here $W(\cdot)$ denotes the Lambert W-function and $a$ is a fit parameter. The corresponding fit to the data is shown in Fig. \ref{length-scales}A, which matches the simulation data very well for low $p$. \\

\textit{Active Fractalytes.}---In mixtures where most particles are attractive (large $p$) and where $Y^*$ is also large, as expected, at the early stages in our simulations we observe that all particles aggregate and form small rigid clusters such that no dilute gas remains. As time evolves, the aggregates move ballistically and self-rotate and once they meet, they typically stick together for very long times, ultimately leading to a (meta)stable macrocluster containing all particles in the system (Fig. \ref{phasediagram}A, red line). Remarkably, this macroscluster is not closely packed but shows a characteristic internal structure comprising numerous holes and gaps (Fig. \ref{snapshots}B, Movie M3 in the SM \cite{supp-info}) leading to a non-integer (fractal) dimension (Fig. \ref{phasediagram}D), as we will further specify below. The fractal dimension decreases as $p$ increases and ultimately transitions for $p \rightarrow 1.0$, where all particles reciprocally attract each other, to a state resembling diffusion limited aggregation (DLA) \cite{Smirnov-PRL-1990}. What is the mechanism leading to active macroclusters with a fractal dimension for $p<1$? The active molecules which form at early times in our simulations typically have a rigid structure as well as a characteristic self-propulsion velocity and self-rotation frequency due to unbalanced nonreciprocal forces. The self-rotations provoke elongated cluster shapes, since the rotating clusters move faster at locations which are far away from the center of rotation, leading to a more efficient recruitment of additional particles which preferably attach to the outer parts of the rotating clusters. At later times, when two (or more) of these elongated motile structures approach each other (Fig. \ref{snapshots}E), they stick together without significantly changing their shapes which ultimately leads to clusters with significant gaps. Note that in contrast to DLA in passive systems, where  individual particles successively bind to a growing structure, here activity promotes the formation of motile rigid building blocks which grow ballistically rather than diffusively and lead to relatively large closed holes. In addition, active fractalytes break up and reform from time to time, so they are dynamic even after complete aggregation, whereas structures forming through DLA are typically stable. To further characterize active fractalytes, we have determined the time-evolution of the number of holes and found that it converges (on average) to a constant value (Fig. \ref{length-scales}B). The oscillations stem from the ongoing dynamics of the fractalytes. That is, active fractalytes do not systematically `heal out' but feature a charateristic distribution of holes and gaps which are persistently renewed by nonreciprocal forces acting on their components. \\

\textit{Fractal dimension.}---To see that the observed gaps and holes indeed lead to a fractal structure, we now analyze the fractal dimension $D_f$. To this end, we sample different cluster sizes and their corresponding radius of gyration $R_g$ which we calculate by using OVITO \cite{Stukowski-ModSimMa-2022}. The connection between the number of particles $M$ belonging to a cluster and $R_g$ is given by \cite{Sander-PRB-1983, Sander-PRL-1981, Jungblut-JPC-2019} 

\begin{equation}
 M = kR_g^{D_f}.
 \label{fit-log}
\end{equation}

By plotting $\mathrm{log}(M)$ over $\mathrm{log}(R_g)$ and fitting a linear function to the data, the slope gives the fractal dimension (inset of Fig. \ref{phasediagram}D). We observe, that the fractal dimension strongly increases if nonreciprocal interactions are present. It rapidly increases from $D_f \approx 1.45$ \cite{Smirnov-PRL-1990, Kolb-PRL-1983} for $p=1.0$ to values around $1.8$ for $p \in [0.85,0.95]$. This steep increase of $D_f$ (Fig. \ref{phasediagram}D) is a clear evidence, that the breaking of the action-reaction symmetry leads to fractals which differ from structures emerging by DLA-like processes which have recently been studied also for active particles \cite{Paoluzzi-PRE-2018}. Note also, that due to cluster-cluster aggregation for $p=1.0$ we obtain $D_f \approx1.45$ rather than $D_f=1.74$ as for ordinary DLA.\\ 

\begin{figure}[h!]
\centering
 \includegraphics[width=0.48\textwidth]{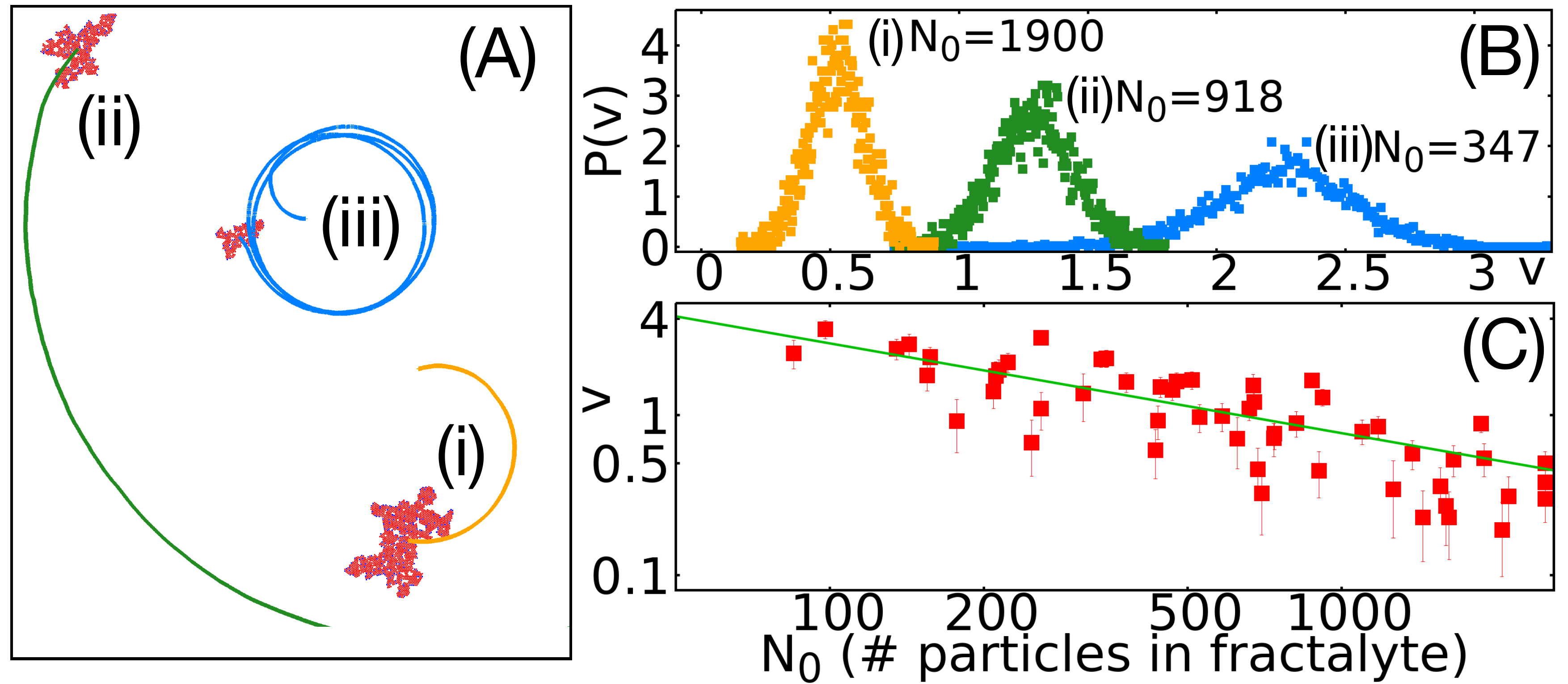}
 \caption{(A) Typical center of mass trajectories of active fractalytes. (B) Speed distribution $P(v)$ for the fractalytes shown in (A). (C) Self-propulsion speed as a function of the fractalyte size. Error bars are the standard deviation of $P(v)$.}
 \label{dynamics}
\end{figure}

\textit{Dynamical properties.}--- To characterize the dynamics of the active fractalytes we calculate the mean-squared displacement (Fig. \ref{phasediagram}E), which reveals a diffusive behavior with an active diffusion coefficient $D_a = \lim \limits_{t^* \to \infty} MSD(t^*)/t^* = 0.6$ both in the regime of dynamical clustering and in the gas-like phase (green and red line, Fig. \ref{phasediagram}E). In both cases self-propulsion is unimportant at late times as most of the emerging aggregates are isotropic (Fig. \ref{snapshots}D). Notably, active fractalytes show a much larger active diffusion coefficient ($D_a\approx 7.3$ for $p=0.95$, blue line, Fig. \ref{phasediagram}E). This is because of the persistent self-propulsion of the fractalytes which increases with $Y^*$. In sharp contrast, for $p=1.0$ (DLA-like regime), the dynamics is subdiffusive with MSD $\sim (t^*)^{0.3}$. In addition, for statistical reasons and for large $N_0$, the self-propulsion speed $v$ of active fractalytes is expected to decrease as $N_0^{-1/2}$, which is consistent with our results (Fig. \ref{dynamics}B,C), where we find $v \sim N_0^{-0.56 \pm 0.06}$. \\

\textit{Phase diagram.}---To summarize our results in a non-equilibrium phase diagram, we calculate the mean cluster size $\langle N_{cl}\rangle$ and the number of single particles $\langle N_s\rangle$ to decide to which phase the system belongs. The criteria are summarized in TABLE \ref{criteria}. In the state diagram (Fig. \ref{phasediagram}C), the transition from the uniform state (blue) to dynamic clustering (green) has the form $Y^* \propto 1/p $, which is plausible since $U_{tot} \propto \Phi_0 p Y^*$ measures the total potential energy in the system. That is, despite the complex phenomenology of the active molecules, which further evolve into dynamic clusters, the onset criterion for their formation can be simply understood from the consideration of the total potential energy which competes with thermal motion. Next to the uniform regime, in the green domain, dynamic clustering shows up, and for even higher $Y^*$ and $p$ active fractalytes (red) occur. These do not necessarily take the form of one macrocluster, but near the transition line, e.g. at $(Y^*,p)=(200,0.5)$ (Movie M4 in the SM \cite{supp-info}), smaller and more dynamic clusters may form, which feature holes and almost no existing gas like phase, which distinguishes them from the dynamic clustering regime.

\begin{table}[!h]
\begin{tabular}[t]{c|c|c|c|c}
                      & \thead{uniform} & \thead{dynamic \\ clusters} & \thead{active \\ fractalytes}& \thead{DLA- \\ like} \\
                      \hline
 $\langle N_{cl}\rangle$               &       $\approx 1$      &             $\ll N$         &   $\rightarrow N$ &    $\rightarrow N$  \\
 $\langle N_s\rangle$            & $\approx N$       &     $>0.01N$           &    $\leq 0.01N$ & $\leq 0.01N$\\
 $D_f$                  &   -                &     -                  & $\approx 1.8$ & $\approx 1.4$ \\
 $D_a$                  &    $\approx 0.6 $     & $\approx 0.6 $         & $\approx 7.3 $    &    - \\
 $MSD \sim (t^*)^b$  &     $b\approx 1$   & $b\approx 1 $      & $b\approx 1$   & $b\approx 0.3$
\end{tabular}
\caption{Characteristica of the different phases.}
\label{criteria}
\end{table}

\textit{Conclusions.}---We found that active molecules can self-organize into dynamic structures which have not yet been reported in the literature. First, our results exemplify a novel route towards dynamic clustering, which exploids nonreciprocal interactions to induce ongoing cluster-fragmentation. Second, we have observed and characterized an entirely new state of active matter in the form of active fractalytes, which feature internal holes, gaps and a fractal dimension. These results offer a route towards designing active materials with optical and mechanical properties which can be controlled via the contained holes which can in turn be designed via the composition of the system.

\ \\
\ \\ 
\vspace{20cm}
\newpage
\newpage

\setcounter{equation}{0}
\setcounter{figure}{0}
\setcounter{table}{0}
\setcounter{page}{1}
\makeatletter
\renewcommand{\theequation}{S\arabic{equation}}
\renewcommand{\thefigure}{S\arabic{figure}}
\renewcommand{\bibnumfmt}[1]{[S#1]}
\renewcommand{\citenumfont}[1]{S#1}





\newpage

\begin{center}
\textbf{Supplemental Material: Collective behavior of active molecules: dynamic clusters, holes and active fractalytes} \\
Sebastian Fehlinger and Benno Liebchen \\
\textit{Technische Universit\"at Darmstadt} \\
(Dated: July 2023)
\end{center}

\textbf{Simulations} \\

The overdamped Langevin equation for particle $i$ ($i=1, \dots, N_a, \dots, N$) reads

\begin{equation}
 \dot{\vbr}_i (t) = \frac{1}{\gamma} \textbf{F}_i + \sqrt{2D_t} \boldsymbol{\eta}_i(t)
\end{equation}

with forces

\begin{equation}
 \textbf{F}_i = -\sum_{\genfrac{}{}{0pt}{} {j=1}{j \neq i}}^{N} \boldsymbol{\nabla}_{\textbf{r}_i} U_S(r_{ij}) - \sum_{\genfrac{}{}{0pt}{} {j=1}{j \neq i}}^{N_a} \boldsymbol{\nabla}_{\textbf{r}_i} U_P(r_{ij}).
\end{equation}

After introducing dimensionless parameters as discussed in the main text, the reduced equations of motions are

\begin{equation}
\begin{split}
 \dot{\textbf{r}}_i^*(t^*) = &-\sum_{\genfrac{}{}{0pt}{} {j=1}{j \neq i}}^{N} \boldsymbol{\nabla}_{\textbf{r}_i^*} U_S^{*}(r_{ij}^*) \\
  &- \sum_{\genfrac{}{}{0pt}{} {j=1}{j \neq i}}^{N_a} \boldsymbol{\nabla}_{\textbf{r}_i^*} U_P^{,*}(r_{ij}^*) + \sqrt{2}\boldsymbol{\eta_i}(t^*).
  \end{split}
\label{langevin_red}
\end{equation}

All particle based simulations are performed with the software package LAMMPS. To handle the overdamped regime, we use the BROWNIAN package, which basically solves the equation 

\begin{equation}
 \mathrm{d}\textbf{r} = \gamma^{-1}\textbf{F}\mathrm{d}t + \sqrt{2D_t}\mathrm{d}\textbf{W},
\end{equation}

where $\mathrm{d}\textbf{W}$ denotes random numbers with zero mean and unit variance. For the implementation of the non-reciprocal interactions, we manually modified the existing pair style for Yukawa interactions. The simulation box has the size $L \times L$, where $L$ is chosen corresponding to a packing fraction of $\Phi_0=N\pi\sigma^2/4L^2=0.1$ with the overall particle number $N$. Further, we apply periodic boundary conditions in both, $x$- and $y$-direction. For all simulations we choose $\mathrm{d}t=10^{-6}$. \\

\textbf{Calculation of the cluster size} \\

For the calculation of the cluster sizes, we use OVITO. We choose a critical distance of $r_{c}^*=1.2$ such that particles with a distance of $r^* \leq r_c^*$ to any particle within a given cluster also belong to this cluster and set the minimum cluster size to $N_{cl}^{\mathrm{min}}=1$, so that even a single particle is counted as one cluster. \\

\textbf{Analyzing the fractal dimension} \\

We calculate the fractal dimension by using

\begin{equation}
  M = kR_g^{D_f},
  \label{df_si}
\end{equation}

where $M$ denotes the number of particles inside a certain cluster and $R_g$ the radius of gyration. This is given by 

\begin{equation}
 R_g = \sqrt{\lambda_1^2 + \lambda_2^2},
\end{equation}

where $\lambda_i$ are the eigenvalues of the gyration tensor

\begin{equation}
 G=\frac{1}{n}\sum_{i=1}^n \begin{pmatrix}
                                  x_ix_i & x_iy_i \\
                                  y_ix_i & y_iy_i \\
                             \end{pmatrix},
\end{equation}

where $n$ is the number of particles belonging to the cluster. We collect the corresponding radii of gyration for clusters with a minimal size of $N_{cl}=2$ particles during the whole simulation and fit equation (\ref{df_si}) to the data to get the fractal dimension. The calculation of $R_g$ is done with OVITO, too. \\

\textbf{Analyzing holes} \\

\begin{figure}[!h]
  \begin{center}
  \includegraphics[width = 0.48\textwidth]{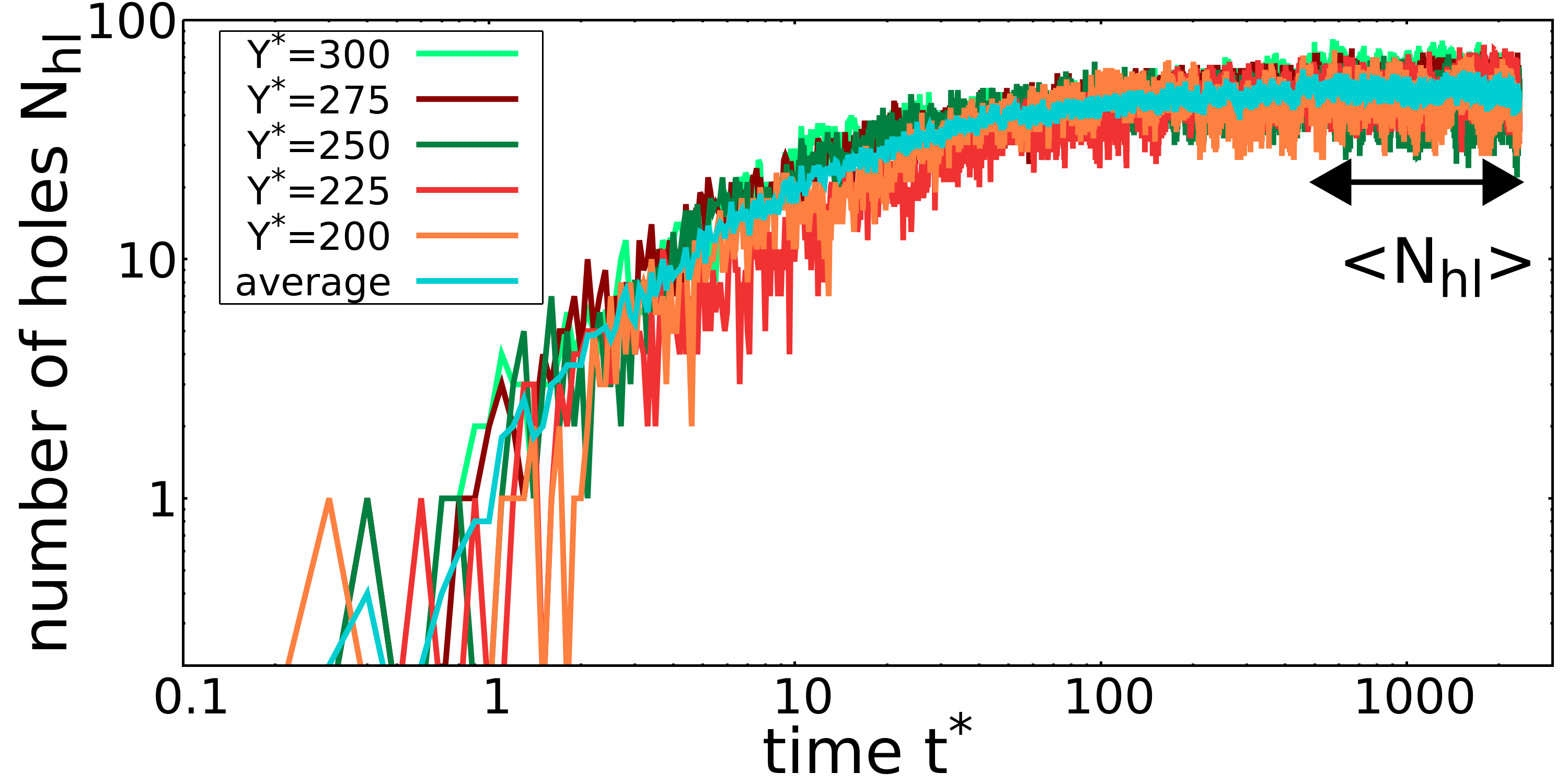}
  \end{center}
 \caption{The number of holes as a function of $t^*$ for systems with $p=0.95$ and different attraction strengths $Y^*$.}
 \label{num-holes}
\end{figure}

\begin{figure*}[]
  \begin{center}
  \includegraphics[width = 0.98\textwidth]{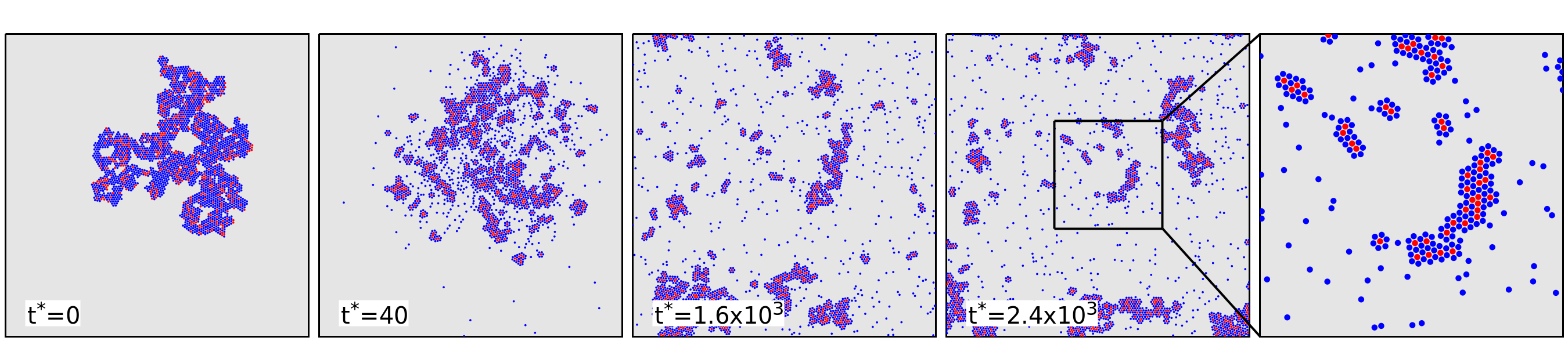}
  \end{center}
 \caption{Simulation snapshots at times $t^*$ showing dynamic clustering for $p=0.25$ with an initial state where all particles belong to one huge macrocluster. Parameters: $N=2.5\times10^3$, $Y^*=300$.}
 \label{ini}
\end{figure*}

\begin{figure}[!h]
  \begin{center}
  \includegraphics[width = 0.42\textwidth]{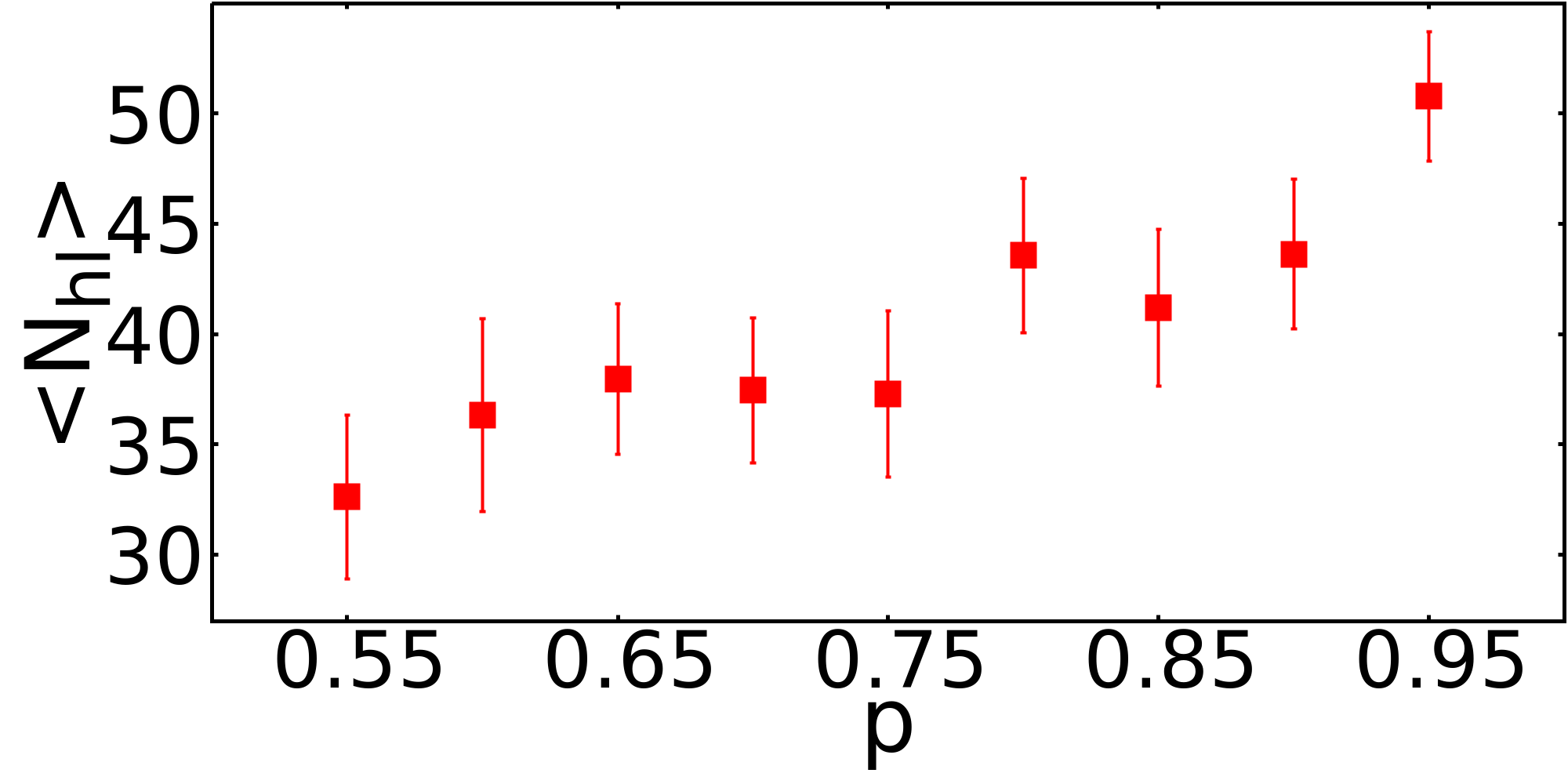}
  \end{center}
 \caption{The average number of holes as a function of $p$.}
 \label{num-holes-1}
\end{figure}

For calculating the number of holes in the active fractalytes, we divide the simulation box into bins with a length of one particle diameter and check, if it contains a particle or not. Then we use the bwlabel command of MATLAB (adopted to account for periodic boundary conditions) to find all neighboring bins, which are empty. Finally, we delete the largest hole corresponding to the environment of the macrocluster(s). Further we introduce a minimal hole size of $N_{hl}^{\mathrm{(min)}}=2$. The number of holes $N_{hl}(t^*)$ is shown in Fig. \ref{num-holes}. After rapidly increasing at early times in our simulations, $N_{hl}(t^*)$ saturates to a characteristic value. The remaining fluctuations in our data stem from the metastable character of the clusters as discussed in the main text. Since $N_{hl}$ does not differ significantly for different values of $Y^*$ and one value of $p$, we average over five values $Y^* \in [200,300]$ (see light blue line in Fig. \ref{num-holes}). To get the average number of holes $\langle N_{hl}\rangle$, we average $N_{hl}(t^*)$ from $t^*=500$ to the end of the simulation. It increases roughly linear with $p$ (Fig. \ref{num-holes-1}), which shows that the density of active fractalytes can be designed via the composition of the underlying colloidal mixture. The curves shown in the main text show a moving average. \\

\textbf{Length scale of dynamical clustering} \\

To understand and predict the characteristic length scale of the dynamic clusters which we observe in our simulations, we compare the passive diffusion coefficient $D_t$ with the Yukawa attractions. Since for small $p$ the clusters are expected to grow up to the point where attractions and diffusion balance, the characteristic length scale $\bar{r}$ is expected to follow the equation

\begin{equation}
 D_t = \frac{Y^*}{\gamma} \frac{\mathrm{e}^{-\mu^* \bar{r}}}{\bar{r}}.
\end{equation}

This transcendental equation can be solved by the Lambert W function with the solution 

\begin{equation}
 \bar{r} = \frac{1}{\mu^*}W\left(\frac{\mu^*}{\alpha}\right)
\end{equation}

where $\alpha=\frac{D_t \gamma}{Y^*}$. We compare this result with our simulations in the main text. \\ 

\textbf{Initial conditions} \\

To explore the role of the initial conditions we now exemplary initialize a large macrocluster with a random distribution of attractive and passive particles inside the cluster. At the beginning of the simulation, the cluster breaks up rapidly (Fig. \ref{ini}) and further evolves towards a final state which is similar to the state which we have encountered for a uniform, random initial distribution: small clusters surrounded by a layer of passive particles and a dilute phase. Notably, here, the mean cluster size does not necessarily  converge to exactly the same value (see dark-green and orange line in Fig. 2, (A) in the main text), but can be lower as for a uniform initial state depending on the precise arrangement of the particles in the initial state. \\ 

\textbf{Movies} \\

The movies show simulations of $N=2.5\times10^3$ particles. Fixed parameters are $\Phi_0=0.1$, $\mu^*=2$, $\epsilon^*=100$. \\

\textit{Movie\_M1:} Dynamic clusters for $(Y^*,p)=(50,0.85)$. \\

\textit{Movie\_M2:} Dynamic clusters for $(Y^*,p)=(300,0.25)$. \\

\textit{Movie\_M3:} Active fractalytes for $(Y^*,p)=(300,0.95)$. \\ 

\textit{Movie\_M4:} Active fractalytes for $(Y^*,p)=(200,0.5)$.

\end{document}